\documentclass[twocolumn,prl,amsmath,amssymb,showpacs,superscriptaddress]{revtex4}
\usepackage{graphicx}
\usepackage{dcolumn}
\usepackage{bm}
\usepackage{latexsym}
\usepackage{amstext}
\usepackage{amsxtra}
\usepackage{color}

\begin{document}

\title{Interlayer superfluidity in bilayer systems of fermionic polar molecules}

\author{A. Pikovski$^{1}$, M. Klawunn$^{1,2}$, G.V. Shlyapnikov$^{3,4}$, and L. Santos$^{}$}
\affiliation{\mbox{$^{}$Institut f\"ur Theoretische Physik, Leibniz Universit\"at Hannover, Appelstr. 2, 30169, Hannover, Germany}\\
\mbox{$^{2}$INO-CNR BEC Center and Dipartimento di Fisica, Universit\`a di Trento, 38123 Povo, Italy}\\
\mbox{$^{3}$Laboratoire de Physique Th{\'e}orique et Mod{\`e}les Statistique, Universit{\'e} Paris Sud, CNRS, 91405 Orsay, France}\\
\mbox{$^{4}$Van der Waals-Zeeman Institute, University of Amsterdam, Valckenierstraat 65/67, 1018 XE Amsterdam, The Netherlands}}

\date{\today}

\begin{abstract}
We consider fermionic polar molecules in a bilayer geometry where they are oriented perpendicularly to the layers, which permits both low inelastic losses and superfluid pairing.
The dipole-dipole interaction between molecules of different layers leads to the emergence of {\it interlayer superfluids}. The superfluid regimes range from BCS-like fermionic superfluidity
with a high $T_c$ to Bose-Einstein (quasi-)condensation of interlayer dimers, thus exhibiting a peculiar BCS-BEC crossover. We show that one can cover the entire crossover regime under 
current experimental conditions.
\end{abstract}

\pacs{67.85.-d,03.75.Ss,74.78.-w}

\maketitle

% INTRODUCTION

Ultracold gases of dipolar particles attract great interest because the 
dipole-dipole interaction drastically 
changes the nature of quantum degenerate regimes compared to ordinary short-range interacting gases~\cite{Baranov2008,Lahaye2009}. This has been demonstrated
in experiments with Bose-condensed chromium atoms which have a magnetic moment of $6\mu_B$ equivalent to an electric dipole moment of 
$0.05$ D~\cite{Lahaye2007,Koch2008,Lahaye2008}. The recent experiments on creating polar molecules in the ground ro-vibrational state~\cite{Ni2008,Deiglmayr2008} and cooling them
towards quantum degeneracy~\cite{Ni2008} have made a breakthrough in the field. For such molecules polarized by an electric field the dipole-dipole interaction 
is several orders of magnitude larger than for atomic magnetic dipoles. This opens fascinating prospects for the observation of new quantum phases~\cite{Baranov2008,Lahaye2009,Pupillo2008,Wang2006,Buchler2007,Jorge2009,Cooper2009}.
The main obstacle is the decay of the system  due to ultracold chemical reactions, such as KRb+KRb$\Rightarrow$K$_2$+Rb$_2$ found in
JILA experiments~\cite{Ospelkaus2010}. These reactions are expected to be suppressed by the intermolecular repulsion in 2D geometries where the molecules are oriented perpendicularly to
the plane of their translational motion~\cite{Ni2010}. 

\begin{figure}%[ht]
%\vspace{2mm}
\begin{center}
\includegraphics[width=0.48\textwidth]{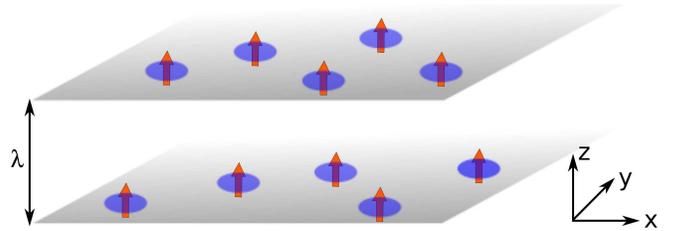}
\caption{Bilayer dipolar system under consideration.}\label{fig:scheme}
\end{center}
%\vspace*{-0.5cm}
\vspace*{-0.6cm}
\end{figure}

In this Letter we consider fermionic polar molecules in a bilayer geometry where the dipoles are oriented perpendicularly to the layers (Fig.~\ref{fig:scheme}), which leads to low
inelastic losses and allows for the possibility of superfluid pairing. The interaction between dipoles of different layers may lead to the emergence of an {\em interlayer superfluid}, that is a superfluid 2D gas where Cooper pairs are formed by fermionic molecules of different layers. We show that the interlayer dipole-dipole interaction provides a higher superfluid transition temperature than that for 2D spin-1/2 fermions with attractive short-range interaction. 
 
Interestingly, an increase in the interlayer dipole-dipole coupling leads to a novel BCS-BEC crossover resembling that studied for atomic fermions near a Feshbach resonance
~\cite{Bloch2008,Giorgini2008}. The reason is that two dipoles belonging to different layers can always form a bound state~\cite{Simon1976}. As long as the binding energy $\epsilon_b$ is much smaller than the Fermi energy $E_F$, or equivalently the size of the interlayer two-body bound state greatly
exceeds the intermolecular spacing in the $\{x,y\}$ plane, the ground state of the system is the BCS-paired interlayer superfluid. Once a reduction of the interlayer spacing $\lambda$ or an increase of the molecular dipole moment $d$ by an electric field make $\epsilon_b >> E_F$, dipolar fermions of different layers form true bound states in real space and the 
ground state is a Bose-condensed system of these composite bosons. We describe this peculiar BCS-BEC crossover and show that interlayer superfluids may be observed for typical parameters of ongoing experiments. Strictly speaking, at a finite temperature $T$ in the thermodynamic limit this is a crossover from a BCS-paired algebraic superfluid to an algebraic bosonic superfluid (quasi-BEC) of dimers.
However, we keep the term BCS-BEC crossover for brevity.

We consider the bilayer system of Fig.~\ref{fig:scheme}, assuming no interlayer hopping. The interaction potential between two dipoles belonging to different 
layers has the form:
\begin{equation}\label{eqV}
V(r)=d^2\frac{r^2-2\lambda^2}{(r^2+\lambda^2)^{5/2}},
\end{equation}
where $r$ is the in-plane separation  between these dipoles. The potential $V(r)$ is attractive for $r<\sqrt{2}\lambda$, and repulsive
at larger distances $r$. It satisfies the relation 
\begin{equation}  \label{relation}
\int  V(r) d^2 r=0, 
\end{equation}
which precludes the ordinary method of finding a bound state in 2D potentials finite at the origin~\cite{LL3}. However, it has been proven that $V(r)$ always has
a bound state~\cite{Simon1976}, at any dimensionless strength $\beta=r_*/\lambda$, with $r_*=md^2/\hbar^2$ being the dipole-dipole length. For $\beta\ll\ 1$ the binding 
energy is exponentially small~\cite{Klawunn2010}: 
\begin{equation}\label{Eb}
\epsilon_b\simeq E_0\exp\left[-8(1-\beta)/\beta^2-(5+2\gamma-2\ln{2})\right],
\end{equation}
where $E_0=\hbar^2/m\lambda^2$, and $\gamma=0.5772$ is the Euler constant. 
One finds numerically that Eq.~(\ref{Eb}) is valid up to $\beta\simeq 1$. Note that the unusual dependence on the interaction, $\epsilon_b \sim \exp (-8/\beta^2)$, 
is a consequence of Eq.~(\ref{relation}).

Dipoles of different layers undergo the 2D $s$-wave scattering from each other in the interlayer potential $V(r)$.
We define the off-shell scattering amplitude as
\begin{equation}       \label{foffshell}
f({\bf k},{\bf k}^{\prime})=(m/\hbar^2)\int \exp(-i{\bf k}'{\bf r})V(r)\psi_{{\bf k}}({\bf r})d^2r,
\end{equation}
where $\psi_{{\bf k}}({\bf r})$ is the true wavefunction of the relative motion with momentum ${\bf k}$.
The potential $V(r)$ shows a slow power law decay $\sim 1/r^3$ at large distances $r$. Therefore, at low relative momenta $k\ll r_*^{-1},\lambda^{-1}$ and $k'\ll r_*^{-1},\lambda^{-1}$ one has two contributions to the scattering amplitude:
the contribution from short distances and the so-called anomalous contribution from distances $r\sim 1/k$~\cite{LL3} obtained using a perturbative approach in $V(r)$. 
The leading short-range and anomalous contributions yield the following $s$-wave part of $f({\bf k},{\bf k}^{\prime})$: 
\begin{equation}\label{ampl-log}
f(k,k')= \frac{2\pi}{\ln(\kappa/k)+i\pi/2}-2\pi kr_*F_1\left(\frac{k^{\prime}}{k}\right), 
\end{equation}
omitting higher order terms. The short-range (logarithmic) contribution is obtained by putting $k'=0$ and proceeding along the lines of the 2D scattering theory~\cite{LL3}. The $k$-dependence of the $s$-wave part of $\psi_{{\bf k}}$ at distances in the interval $r_*,\lambda\ll r\ll k^{-1}$ is given by a factor $[\ln(\kappa/k)+i\pi/2]^{-1}$, where $\kappa$ depends on the behavior of $V(r)$ at small $r$ and in the presence of the weakly bound state we have $\kappa=\sqrt{m\epsilon_b}/\hbar$~\cite{LL3}. The anomalous term comes from distances where the motion is almost free. We then have $F_1(x)=(x^2/2)F(1/2,1/2,2,x^2)+F(1/2,-1/2,1,x^2)$, where $F$ is the hypegeometrical function, so $F_1(1)=4/\pi$. This is valid for $k^{\prime}<k$, for $k^{\prime}>k$ one should interchange $k^{\prime}$ and $k$. A detailed derivation of $f(k,k')$, including $k^2$-terms, will be given elsewhere.
  
The anomalous term in Eq.~(\ref{ampl-log}) corresponds to attraction and so does the logarithmic term if $\kappa\ll k$, i.e. if the collision energy is much larger than $\epsilon_b$. Thus, both the short-range and anomalous contribution may lead to superfluid interlayer pairing. Note that for short-range potentials, like all interatomic potentials decaying as $1/r^6$,
only the logarithmic term is present in Eq.~(\ref{ampl-log}). We will show that the anomalous scattering drastically influences the superfluid pairing. The inlayer dipole-dipole interaction is repulsive and it simply renormalizes the chemical potential. This is valid as long as the inlayer repulsion is sufficiently weak to exclude crystallization~\cite{Buchler2007,Jorge2009}.

Treating molecules of the first and second layers as spin-up and spin-down fermions our problem is mapped onto spin-$1/2$ fermions with a peculiar interaction potential. 
For a weak interlayer attractive interaction we use the BCS approach and obtain the standard gap equation for the momentum-space order parameter: 
%The order parameter is introduced as
%\begin{equation}\label{Def-gap}
%\Delta({\bf k})=\int d^2 r \langle \hat\Psi_1({\bf r})\hat\Psi_2(0) \rangle 
%V(r) \exp(i{\bf kr}),
%\end{equation} 
%with $\hat\Psi_i$ being the field operators of the atoms in layer $i$, and we  
\begin{equation}\label{gap-equ}
\Delta({\bf k})=-\int \frac{dk'^2}{(2 \pi)^2}\frac{V({\bf k}-{\bf k}')\Delta({\bf k}')}{2\epsilon_{k'}}\tanh \left ( \frac{\epsilon_{k'}}{2T} \right ),
\end{equation}
where $\epsilon_k=\sqrt{\left(E_{k}-\mu\right)^2+|\Delta(k)|^2}$ is the gapped dispersion relation, $\mu$ is the chemical potential, and $E_k=\hbar^2k^2/2m$.
We rewrite Eq.~(\ref{gap-equ}) expressing the Fourier component of the interaction potential, $V({\bf k}-{\bf k}')$, through the off-shell scattering amplitude ~\cite{AGD}. Assuming that 
the $s$-wave interaction is the leading channel of superfluid pairing the renormalized gap equation reads:
\begin{eqnarray}\label{gap-equ2}
&&\Delta(k)=-\frac{\hbar^2}{2m} \int \frac{d^2k'}{(2\pi)^2} f(k,k') \Delta(k')  \nonumber \\
&&\times\left\{\frac{\tanh(\epsilon_{k'}/2T)}{\epsilon_{k'}}-\frac{1}{E_{k'}-E_k-i0} \right\}.
\end{eqnarray}

As long as the interaction is really weak and $\mu\simeq E_F$~\cite{footnote-Fermi}, Eq.~(\ref{gap-equ2}) may be employed for calculating $\Delta(k)$ and the superfluid transition temperature. As known, in 2D the transition from the normal to superfluid state is of the Kosterlitz-Thouless type. However, in the BCS limit the Kosterlitz-Thouless 
transition temperature is very close to the critical temperature $T_c$ given by the BCS gap equation ~\cite{Miyake1983}. Using Eq.~(\ref{gap-equ2}) we 
obtain the relation between $T_c$ and the order parameter on the Fermi surface at $T=0$, $\Delta_0(k_F)$, which is the same as in 3D~\cite{Landau9}:  
\begin{equation}   \label{TcDelta0}
T_c=(e^{\gamma}/\pi)\Delta_0(k_F).
\end{equation} 

When the short-range logarithmic contribution to Eq.~(\ref{ampl-log}) dominates, the anomalous term can be omitted. This is in particular the case for 
$\beta$ approaching unity and sufficiently small values of $k_F\lambda$. Then, using Eq.~(\ref{gap-equ2}) we recover the well-known results~\cite{Miyake1983,Randeria1990}:
\begin{equation}  \label{T-anal}
\Delta_0(k_F)=\sqrt{2E_F\epsilon_b};\,\,\,\,\,\,T_c=(e^{\gamma}/\pi)\sqrt{2E_F\epsilon_b}.
\end{equation}
Note that we do not include here the second order Gor'kov--Melik-Barkhudarov corrections. They decrease both $\Delta_0(k_F)$ and $T_c$ by a factor of $e$ ~\cite{Petrov2003}, 
but Eq.~(\ref{TcDelta0}) remains valid. A detailed analysis of the BCS limit up to the second order will be given elsewhere.

For $\beta<1$ the anomalous scattering dominates, at least for not very low 
$k_F$. 
Then the logarithmic term in Eq.~(\ref{ampl-log}) reduces to $-\pi\beta^2/2$, and it is necessary to include quadratic terms in $k$. In this case the scattering amplitude can be calculated using the second order Born approximation.
The expression for the off-shell amplitude is cumbersome. The on-shell amplitude is $f(k,k)\!\!=\!\!f'/(1\!+\!if'/4)$, where $f'(k)$ is real and given by 
\begin{eqnarray}     \label{ampl-born2}
\!\!\!\!f'\!\!=\!\!-8kr_*\!+\!\frac{4\pi(kr_*)^2}{\beta}\!-\!\frac{\pi\beta^2}{2}\!+\!3\pi (kr_*\!)^2\!\ln\!\left[\!\frac{k\lambda}{2}e^{\gamma+\frac{23}{12}}\!\right]\!\!,\!
\end{eqnarray}
where the first term is dominant and it follows from the second term in Eq.~(\ref{ampl-log}) at $k'=k$.  

\begin{figure}%[b]
%\vspace{0.5cm}
%\vspace*{-0.2cm}
\begin{center}
\includegraphics[width=0.28\textwidth,angle=270]{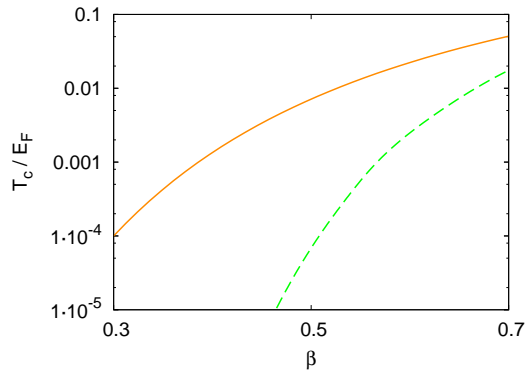}
%\vspace*{-0.2cm}
\caption{Critical temperature $T_c$
as a function of the dipole-dipole strength $\beta$ for $k_F \lambda=0.5$.
The numerical solution (solid) is higher by at least an order of magnitude than the result of Eq. (\ref{T-anal}) (dashed).}
\label{fig:Tcrand}
\end{center}
\vspace*{-0.6cm}
\end{figure}

We now use Eq.~(\ref{gap-equ2}) to calculate $T_c$. For $T\rightarrow T_c$, we set $\Delta(k')=0$ in the dispersion relation which becomes $\epsilon_{k'}=|E_{k'}-E_F|$. 
The main contribution to the integral over $dk'$ comes from the region near the Fermi surface, where we put
$k'=k_F$ in the arguments of $\Delta$ and $f$, and taking $k=k_F$ use $f(k_F,k_F)$ from Eq.~(\ref{ampl-born2}). For the rest of the integration it is sufficient to use $f(k_F,k')$ given by the second term of Eq.~(\ref{ampl-log}) and employ the relation $\Delta(k)\simeq\Delta(k_F)f(k,k_F)/f(k_F,k_F)$ following from Eq.~(\ref{gap-equ2}). After a straightforward algebra we then find:
\begin{equation}\label{Tc-born2}
T_c=0.1E_F\left(\frac{E_0}{E_F}\right)^{\!0.46}\!\!\!\!\!\!\exp\left\{-\frac{\pi}{4k_Fr_*}G(k_F\lambda,\beta)\right\},
\end{equation}
where $G(x,y)=(1-\pi x/2+\pi y/16x)^{-1}$. The validity of Eq.~(\ref{Tc-born2}) requires $1\gg k_F\lambda\gg\pi\beta/16$.
One easily checks that Eq.~(\ref{Tc-born2}) gives a significantly higher $T_c$ than that given by Eq.~(\ref{T-anal}).
The numerical solution of Eq.~(\ref{gap-equ2}) also confirms this conclusion 
%~\cite{numerics} 
for $k_F\lambda$ approaching unity. Fig.~\ref{fig:Tcrand} shows 
that $T_c$ strongly departs from Eq.~(\ref{T-anal}) for small $\beta$, 
an anomalous behavior stemming from the long-range character of the interlayer interaction.

For sufficiently strong interactions $\mu$ deviates from $E_F$, and Eq.~(\ref{gap-equ2}) should be complemented by the number equation~\cite{Eagles1969,Leggett1980} for the 2D density $n$ in one layer:
\begin{equation}\label{N-equ}
n=\frac{1}{2}\int \frac{d^2 k}{(2\pi)^2}\left\{1-\frac{E_k-\mu}{\epsilon_k}\tanh\left(\frac{\epsilon_k}{2 T}\right)\right\}.
\end{equation}
We found numerically $\mu$ and $\Delta(k)$ from the self-consistent solution of Eqs.~(\ref{gap-equ2}) and~(\ref{N-equ}). Alternatively, we used Eq.~(\ref{gap-equ}) together with~(\ref{N-equ}), which is adequate since the potential $V(r)$ is finite and strongly bounded at small $r$.

\begin{figure}%[ht]
\vspace{-0.2cm}
\begin{center}
\vspace*{-0.7cm}
\includegraphics[width=0.35\textwidth,angle=270]{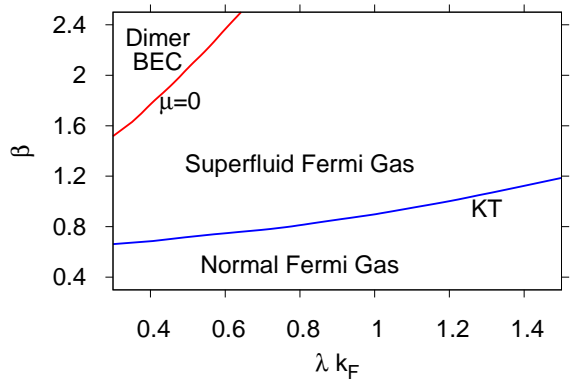}
\vspace*{-0.4cm}
\caption{Phase diagram for $T=0.05 E_F$, obtained from Eqs.~(\ref{gap-equ2}) and (\ref{N-equ}). The curves indicate the Kosterlitz-Thouless~(KT) transition 
and the $\mu=0$ line.}
\label{fig:BCSBEC}
\end{center}
\vspace*{-0.6cm}
\end{figure}

This approach provides a qualitative description of the strongly interacting regime ~\cite{Eagles1969,Leggett1980}. An increase of $\epsilon_b$ by e.g. increasing $\beta$ leads to 
bound interlayer dimers when $\epsilon_b$ becomes much larger than $E_F$ and the chemical potential for the fermionic molecules is $\mu\simeq -\epsilon_b/2$, in contrast to $\mu\simeq E_F$ in the BCS regime. These composite bosons condense and we thus have a BCS-BEC crossover. 
An approximate crossover line is marked by the condition 
$\mu=0$~\cite{Leggett1980}~(Fig.~\ref{fig:BCSBEC}). At sufficiently low $T$, well above this line a dimer (quasi)BEC occurs whereas well below the line the system is a Fermi gas which 
is superfluid or normal, depending on $T$ and density (Fig.~\ref{fig:BCSBEC}).  
 
\begin{figure}%[b]
%\vspace{0.5cm}
\begin{center}
\includegraphics[width=0.28\textwidth,angle=270]{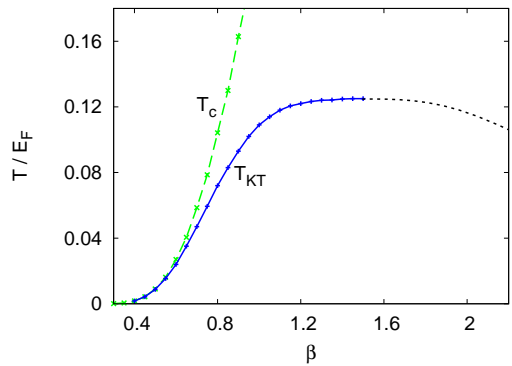}
%\vspace*{-0.2cm}
\caption{Kosterlitz-Thouless transition temperature $T_{KT}$ and the critical BCS temperature $T_c$
versus the dipole-dipole strength $\beta$ for $k_F \lambda = 0.5$.}
\label{fig:TKT}
\end{center}
\vspace*{-0.8cm}
\end{figure}

For strong interactions, $T_c$ calculated from Eqs.~(\ref{gap-equ2}) and~(\ref{N-equ}) 
cannot be interpreted as the critical temperature for the onset of superfluidity. 
Instead, it corresponds to the temperature of pair dissociation~\cite{SadeMelo1993}. 
The temperature of the Kosterlitz-Thouless transition, $T_{KT}$, below which the system is superfluid
satisfies the equation~\cite{Nelson1977}:
\begin{equation} 
% k_B T_{KT}=\frac{1}{2}\pi\hbar^2 \frac{\rho_s(T_{KT})}{M^2},
T_{KT}=\pi\hbar^2 \rho_s(T_{KT})/2 M^2,
\end{equation}
where $M=2m$ is the dimer (Cooper-pair) mass, and $\rho_s$ is the superfluid mass density just below $T_{KT}$, which may be determined from our mean-field equations using 
the known expression for the normal density~\cite{Landau9}.
%\begin{equation}\label{eq-rho}
% \rho_n=\frac{\hbar^2}{2\pi}\int_0^\infty \frac{-\partial n_F}{\partial \epsilon_k} k^3 dk
%\end{equation}
%and hence $\rho_s=2m n_{2d}-\rho_n$, with $n_F(\epsilon)=(1+e^{\epsilon/T})^{-1}$.
In Fig.~\ref{fig:TKT} we depict $T_{KT}$ and $T_c$ versus $\beta$ for $k_F\lambda=0.5$. 
In the BCS regime we retrieve $T_{KT} \approx T_c$ and see that the ratio $T_c/E_F$ can reach $0.04$. For strong interactions where $\Delta(k_F)$ is a sizable fraction of $E_F$ we obtain $T_{KT}=0.125 E_F$ (cf.~\cite{Botelho2006}). In the intermediate regime $T_{KT}$ interpolates smoothly between 
these limits ~(see Fig.~\ref{fig:TKT}). 
On the BEC side of the crossover we take into account a noticeable normal fraction, which makes $T_{KT}$ lower. We obtain that $T_{KT}\simeq 0.1E_F$
for $\beta\simeq 2.2$ and very slowly decreases with increasing $\beta$ (decreasing the interaction), in agreement with the result for bosons ~\cite{Prokofiev-Svistunov2001}.

\newpage
In conclusion, we have shown that bilayer systems of fermionic polar molecules which are expected to have low inelastic losses, at the same time may allow the observation
of interesting regimes of interlayer superfluidity. These regimes range from fermionic BCS-like superfluidity with a relatively high $T_c$ and Cooper pairs formed by molecules of different layers,
to quasiBEC of interlayer dimers, thus exhibiting a peculiar BCS-BEC crossover. For example, by making the interlayer spacing $\lambda\simeq 250$ nm one achieves $k_F\lambda\simeq 2$
for KRb and LiCs molecules at densities $n\simeq 5\,10^8$ cm$^{-2}$ corresponding to $E_F\simeq 110$ nK. Then, varying the LiCs dipole moment $d$ from $0.35$ to $1.3$ D by increasing 
the electric field to about $1$ kV/cm, one obtains $\beta$ ranging from $1$ to $14$ and covers the entire crossover regime, with $T_{KT}$
of a few nanokelvin. For KRb molecules the strongly interacting regime can be reached for the presently achieved $d\simeq 0.2$ D ~\cite{Ni2010} by putting a shallow in-plane optical lattice
and getting $\beta>1$ due to an increase in the effective mass of molecules. 

Our results open exciting perspectives for future studies. Imbalanced Fermi mixtures 
may be studied by preparing layers with different chemical potentials (effective magnetic field) or 
with different densities. An increase in the dipole-dipole interaction may lead to 
in-plane Wigner-like dimer crystallization, and perhaps opens routes towards a supersolid dimer gas.

We are grateful to T. Vekua and D. Jin for helpful discussions. A. P. was supported by the DFG (QUEST Cluster). 
We acknowledge support from the ESF (EuroQuasar and EuroQuam programs). G.S. acknowledges support from the IFRAF Institute,
from ANR (grant 08-BLAN0165), and from the Dutch Foundation FOM.

{\it Note added:} After the completion of this work we learned of a recent
related work of Potter et al.~\cite{Potter2010}, where interlayer
dimerization and superfluidity in multistacks of polar Fermi molecules have
been considered.


\begin{thebibliography}{99}
\bibitem{footnote-Fermi} Since the most important momenta are those near the Fermi surface, $k,k'\simeq k_F$, this requires ${\rm Re}f(k_F,k_F)<0$ and $|f(k_F,k_F)|/2\pi\ll 1$.
\end{thebibliography}
\end{document}